\documentclass[prd,tightenlines,nofootinbib,amsfonts,amssymb,amsmath,color,11pt]{revtex4}

\usepackage{graphicx}	
\usepackage{color}

\usepackage{amsfonts}
\usepackage{amsmath}
\usepackage{amssymb}
\usepackage{hyperref}

\def\be{\begin{equation}} 
\def\ee{\end{equation}}
\def\bea{\begin{eqnarray}}
\def\eea{\end{eqnarray}}

\def\V{\mathcal{V}}

\def\Mpl{M_{Pl}}

\begin{document}

\title{Prospects for Primordial Gravitational Waves in String Inflation\footnote{\it Essay written for the Gravity Research Foundation 2016 Awards for Essays on
Gravitation}}
\author{Susha L. Parameswaran$^a$} 
\email{Susha.Parameswaran@liverpool.ac.uk}
\author{Ivonne Zavala$^b$} 
\email{e.i.zavalacarrasco@swansea.ac.uk}
\affiliation{$^a$Department of Mathematical Sciences, University of Liverpool,
Liverpool, L69 7ZL, UK }
\affiliation{ $^b$Department of Physics, Swansea University, Swansea, SA2 8PP, UK}

\begin{abstract}
Assuming that the early universe had (i) a description using perturbative string theory and its field theory limit (ii) an epoch of slow-roll inflation within a four-dimensional effective field theory and a hierarchy of scales $M_{inf} <  m_{kk} < m_s \lesssim M_{pl}$ that keeps the latter under control, we derive an upper bound on the amplitude of primordial gravitational waves.  The bound is very sensitive to mild changes in numerical coefficients and the expansion parameters.  For example, allowing couplings and mass-squared hierarchies $\lesssim 0.2$, implies $r \lesssim 0.05$, but asking more safely for hierarchies $\lesssim 0.1$, the bound becomes $r \lesssim 10^{-6}$.  Moreover, large volumes -- typically used in string models to keep backreaction and moduli stabilisation under control -- drive $r$ down.  Consequently, any detection of inflationary gravitational waves would present an interesting but difficult challenge for string theory.

\bigskip
\bigskip
\end{abstract}

\maketitle

\enlargethispage{\baselineskip}
\enlargethispage{\baselineskip}

 The recent first detection of gravitational waves in the fabric of spacetime \cite{LIGO} opens up an entirely new and powerful way to study our Universe.  One very exciting but challenging prospect, is the measurement of primordial gravitational waves (PGWs) produced in the very early universe.  

During the inflationary epoch \cite{Guth,Linde}\footnote{See \cite{ekpyrotic} and \cite{branegas} for alternative models of the early universe.}, the quantum fluctuations in the inflaton and metric tensor fields 
stretched to observables scales \cite{MCh}, setting up the initial conditions for structure growth.  These density (scalar)  perturbations and gravitational (tensor)  waves are measured in the cosmic microwave background (CMB) emitted during the epoch of recombination.  In particular, whilst the dominant contribution to the CMB temperature anisotropies is from density perturbations, gravitational waves lead to B-modes in the CMB polarisation \cite{Zalda,Kamion}.  These B-modes are being searched for by a wide range of ground-based, balloon and satellite experiments (see \cite{Paolo} for a review), with current bounds on the tensor-to-scalar ratio, $r$, from BICEP/Keck set at $r < 0.07$ \cite{BICEP15} and sensitivities from future satellites such as PRISM expected to reach $r \sim 10^{-4}$ \cite{PRISM}.  Moreover, searches for B-modes in the  lensing distortions of the 21 cm radiation emitted by hydrogen atoms during the reionisation epoch could potentially measure primordial gravitational waves as small as $r \sim 10^{-9}$ \cite{21cm}.  Another promising possibility is the direct detection of PGWs with laser interferometry (see \cite{Chongchitnan:2006pe} for a review).

Although PGWs are a robust prediction of inflation, their amplitude depends on the inflationary model, and in particular the inflationary energy scale.  
Analysis of the primordial scalar and tensor perturbations, together with the measured amplitude of scalar perturbations \cite{Planck15}, gives
the following relation  between PGWs and the energy density $\rho_{inf}$ during  inflation:
\be
M_{inf} \equiv \rho_{inf}^{1/4} \approx \left(\frac{r}{0.1}\right)^{1/4} \times 1.8 \times 10^{16} \textrm{GeV}\,. \label{E:Minf}
\ee
This relation assumes a single, canonically normalised scalar field, minimally coupled to gravity and slowly rolling down an almost flat potential during the scales probed by the CMB.  It tells us something very important: because the dependence on $r$ is very weak, the inflationary scale is close to the GUT scale \eqref{E:Minf} for values of $r$ as small as $r\sim 10^{-5}$!
 
Moreover, the tensor-to-scalar ratio is also  related to the evolution of the inflaton field.
Assuming conservatively that $r$ remains approximately constant during the inflationary period probed by the CMB, the inflaton must satisfy the so-called Lyth bound (taking into account that $r$ does not remain constant the bound is much stronger \cite{GBRSZ}) \cite{Lyth,Lotfi}:
\be\label{LB}
\frac{\Delta\phi}{\Mpl} \gtrsim 2 \times \left(\frac{r}{0.01}\right)^{1/2} \,.
\ee
Therefore, it is clear that an observation of primordial gravitational waves with $r \sim 10^{-1}-10^{-2}$ would fix the scale of  inflation to be around the GUT scale and the inflaton field range to be super-Planckian.  In other words, an observation of primordial gravitational waves would imply that inflation is highly sensitive to quantum gravity effects.

The intrinsic sensitivity of inflation -- and especially large field inflation with observable gravitational waves -- to Planck scale physics has motivated a vast amount of work searching for viable models within string theory.  Natural inflation \cite{NI}, axion monodromy \cite{AM1,AM2} and fibre inflation \cite{FI} are the leading candidates for large field inflation in string theory, with much interest generated from their predictions of observable gravitational waves.  The main focus in developing these models has been a string theoretic derivation of the inflaton potential in regimes where backreaction and moduli stabilisation are well under control.

However, any string model of inflation has to feature the following hierarchy of scales \cite{BM, Mazumdar:2014qea, KPZ, Burgess:2016owb}: 
\be
\Mpl \gtrsim m_s \gtrsim m_{kk} > M_{inf}\,, \label{E:hierarchy}
\ee
where $m_s = \sqrt{\alpha'}$ is the mass of the first excited string states and $m_{kk}\sim 1/L$ the lightest Kaluza-Klein mass.  This is necessary in order to ensure that the four-dimensional effective field theory (4D EFT) description is valid throughout the inflationary epoch.  Indeed, during inflation the inflaton must carry energy well below the UV cutoff of the 4D EFT, which is used to derive inflation and its observables.  
In particular, if the inflaton carries energy greater than the compactification scale, $M_{inf} \gtrsim m_{kk}$, then physics is extra-dimensional, and if the inflaton energy is comparable to the string scale, $M_{inf} \sim m_s$, then one  could not even use an EFT description to derive an inflationary background and its observables.  This would be surprising given the remarkable success that inflation -- as a particular class of 4D EFTs -- has when compared to observations.  One immediately sees that for  observable gravitational waves, where $M_{inf} \sim 10^{16}$GeV, there is not much room for these hierarchies to be achieved.  Note morever that, when model building, further hierarchies might be necessary. In the simplest effective single-field models of inflation, $m_{kk} > m_{mod} > M_{inf}$, with $m_{mod}$ the lightest (time-dependent) mass of the (hundreds of) moduli that have been truncated, would ensure that no large moduli kinetic energies are sourced as the inflaton rolls over large field and potential energy ranges. Otherwise the full multifield potential would have to be used to correctly identify a slow-roll trajectory and compute observables\footnote{See \cite{Anna,flattening,NGs} for some examples of how heavy moduli can affect inflation.}.  Also, it has to be checked that the inflationary energy does not overcome barriers in the moduli potential and lead to destabilisation \cite{KL}.
 
The relationship between the string and the Planck scales in a generic compactification of perturbative string theory is obtained by dimensional reduction of the ten-dimensional Einstein-Hilbert term, in the supergravity limit, as: 
\be
M_s= \Mpl \frac{g_s}{\sqrt{4\pi \V_6}} \label{E:MsMpl}
\ee
where $g_s$ is the string coupling, $M_s = 1/\ell_s$ is the string scale 
(with $\alpha' = \ell_s^2/(2\pi)^2$) and $\V_6$ is the possibly warped, string-frame volume of the six extra dimensions in string units. Then, using \eqref{E:Minf} and \eqref{E:MsMpl}, the tensor-to-scalar ratio can be written in terms of the hierarchies in \eqref{E:hierarchy}:
\be
r= 3.1 \times 10^8 \left(\frac{M_{inf}}{m_{kk}}\right)^4 \left(\frac{m_{kk}}{m_{s}}\right)^4 \left(\frac{g_s}{\sqrt{\V_6}}\right)^4 \,. \label{E:rsup}
\ee
Note that $r$ is very sensitive to mild changes in $g_s$, volumes/curvatures (including numerical 2$\pi$ factors) and mass hierarchies, due to the fourth powers in \eqref{E:rsup}.
Assuming $\V_6 l_s^6\sim \beta L^6$ and asking for mass-squared hierarchies\footnote{Note that corrections to an EFT  with cutoff $\Lambda$ usually go as $M^2/\Lambda^2$.} and string coupling to be less than some small number, $\delta$, this implies $\V_6 > \beta/((2\pi)^6 \delta^3)$ and a simple bound\footnote{See \cite{Mazumdar:2014qea} for an earlier discussion of how the 4D EFT for inflation gives bounds relating $r$, $g_s$ annd $\V_6$.} on $r$:
\be
r < 3.1 \times 10^8 \, \frac{(2\pi)^{12}}{\beta^2}\delta^{14}. \label{E:rbound}
\ee
For example, assuming a torus with $\beta \sim (2\pi)^6$, and asking conservatively that $\delta \lesssim 0.1$ (so $M_{inf} \sim 0.3 \, m_{kk} \sim 0.3 \,m_s$) gives:
\be
r \lesssim 3.1 \times 10^{-6} \label{E:bound1}\,.
\ee
Relaxing the couplings and mass hierarchies to $ \delta \lesssim 0.2$ (so $M_{inf} \sim 0.45\, m_{kk} \sim 0.45 \,m_{s}$) allows:
\be%
r \lesssim 0.05 \label{E:bound2}\,.
\ee
We see that any observable $r$ from string theory will depend sensitively on explicit numerical factors and, moreover, be right at the limits of validity of the 4D EFT.  

\enlargethispage{\baselineskip}
\enlargethispage{\baselineskip}
Note that in explicit, controlled string constructions $r$ tends to be small.  For example, in axion monodromy models, long warped throats within throats are used to prevent brane anti-brane annihilation and suppress brane backreaction \cite{AM1,AM2}.  The large internal volume then drives the string scale down, and thus -- via \eqref{E:hierarchy} and \eqref{E:Minf} -- also $M_{inf}$ and $r$ down.  Similarly, fibre inflation is realised using the LARGE volume scenario, where internal volumes are large order to keep moduli stabilisation under control \cite{FI,Burgess:2016owb}.  Any claim of large $r$ in string theory must examine carefully whether numerical factors in explicit models allow the required hierarchy \eqref{E:hierarchy} to be achieved, and check that corrections to the 4D EFT are sufficiently suppressed.  For example, fiber inflation\footnote{See \cite{Burgess:2016owb} for a further discussion on the robustness of fiber inflation.} might achieve $r  \sim 10^{-3}$ with $\V_6 \sim 125$ and $\delta\sim 0.2$, plausibly at the limits of control.

Let us now comment on the robustness of the bounds obtained above. First, one  could try to evade the bounds by  going to strong coupling $g_s>1$ or strong curvatures $L/\ell_s <1$, to drive $m_s$, $m_{kk}$ to higher values. But in such a case, eq.~\eqref{E:MsMpl} would not be valid.  In this case, one  could always perform a duality transformation to an equivalent weak coupling, weak curvature description and return to the bound \eqref{E:bound1} with the same conclusions.

Also, the relationship between $r$ and $M_{inf}$ in \eqref{E:Minf}, and the Lyth bound \eqref{LB},  assume that inflation was driven by a single, canonically normalised inflaton field slowly rolling down a flat potential.  One may wonder, therefore, if having more fields could help evade the bounds derived above.  Additional scalar fields provide an extra source of primordial scalar perturbations, but do not affect the gravitational waves.  It follows that \eqref{E:Minf} remains unchanged \cite{SasakiStewart, Wands}.  Alternatively, we may consider inflation driven by nontrivial kinetic effects rather than a flat potential (``k-inflation'' \cite{kinflation}).  The non-canonical kinetic terms change the speed of sound for the scalar perturbations, but again \eqref{E:Minf} remains unchanged \cite{Garriga:1999vw}.  On the other hand, the bounds may not apply to direct detection of primordial gravitational waves, as these waves would correspond to scales vastly different to those probed by the CMB. 

\enlargethispage{\baselineskip}
\enlargethispage{\baselineskip}

Now in deriving the bounds we have on the other hand assumed  {\it (i)} perturbative string theory and its supergravity limit as a good description of the early Universe {\it (ii)} inflation within a four-dimensional effective field theory, with a hierarchy of scales that controls the latter approximation.   A detection of inflationary gravitational waves in the near future $r \sim 10^{-1} - 10^{-3}$ would therefore suggest that the early Universe was very much at the limits of string perturbation theory and the supergravity limit, and moreover at the limits of the validity of the 4D EFT.  Whilst this would make convincing string realisations of inflation even more challenging, where proper attention must be paid to the required hierarchy of scales \eqref{E:hierarchy}, it would be extremely exciting.  Quantum gravity would have left its imprints in the sky.

\acknowledgements 

{We are grateful to Cliff Burgess, Michele Cicoli, Shanta de Alwis and Fernando Quevedo for helpful conversations on this topic, and valuable comments on the first version of this essay.  We also thank Ralph Blumenhagen, Martin Gorbahn, Renata Kallosh and Clemens Wieck for interesting discussions.  SLP is supported by a Marie Curie Intra European Fellowship within the 7th European Community Framework Programme.

\bibliography{refs}

\providecommand{\href}[2]{#2}\begingroup\raggedright\begin{thebibliography}{10}

\bibitem{LIGO}
{\bf Virgo, LIGO Scientific} Collaboration, B.~P. Abbott {\em et al.},
  ``{Observation of Gravitational Waves from a Binary Black Hole Merger}'',
  \href{http://dx.doi.org/10.1103/PhysRevLett.116.061102}{{\em Phys. Rev.
  Lett.} {\bf 116} (2016) no.~6, 061102},
\href{http://arxiv.org/abs/1602.03837}{{\tt arXiv:1602.03837 [gr-qc]}}.

\bibitem{Guth}
A.~H. Guth, ``{The Inflationary Universe: A Possible Solution to the Horizon
  and Flatness Problems}'',
\href{http://dx.doi.org/10.1103/PhysRevD.23.347}{{\em Phys.Rev.} {\bf D23}
  (1981)  347--356}.

\bibitem{Linde}
A.~D. Linde, ``{A New Inflationary Universe Scenario: A Possible Solution of
  the Horizon, Flatness, Homogeneity, Isotropy and Primordial Monopole
  Problems}'',
\href{http://dx.doi.org/10.1016/0370-2693(82)91219-9}{{\em Phys.Lett.} {\bf
  B108} (1982)  389--393}.

\bibitem{ekpyrotic}
A.~Ijjas and P.~J. Steinhardt, ``{Implications of Planck2015 for inflationary,
  ekpyrotic and anamorphic bouncing cosmologies}'',
  \href{http://dx.doi.org/10.1088/0264-9381/33/4/044001}{{\em Class. Quant.
  Grav.} {\bf 33} (2016) no.~4, 044001},
\href{http://arxiv.org/abs/1512.09010}{{\tt arXiv:1512.09010 [astro-ph.CO]}}.

\bibitem{branegas}
R.~H. Brandenberger, ``{String Gas Cosmology after Planck}'',
  \href{http://dx.doi.org/10.1088/0264-9381/32/23/234002}{{\em Class. Quant.
  Grav.} {\bf 32} (2015) no.~23, 234002},
\href{http://arxiv.org/abs/1505.02381}{{\tt arXiv:1505.02381 [hep-th]}}.

\bibitem{MCh}
V.~F. Mukhanov and G.~V. Chibisov, ``{Quantum Fluctuation and Nonsingular
  Universe. (In Russian)}'',
{\em JETP Lett.} {\bf 33} (1981)  532--535.

\bibitem{Zalda}
M.~Zaldarriaga and U.~Seljak, ``{An all sky analysis of polarization in the
  microwave background}'',
  \href{http://dx.doi.org/10.1103/PhysRevD.55.1830}{{\em Phys. Rev.} {\bf D55}
  (1997)  1830--1840},
\href{http://arxiv.org/abs/astro-ph/9609170}{{\tt arXiv:astro-ph/9609170
  [astro-ph]}}.

\bibitem{Kamion}
M.~Kamionkowski, A.~Kosowsky, and A.~Stebbins, ``{Statistics of cosmic
  microwave background polarization}'',
  \href{http://dx.doi.org/10.1103/PhysRevD.55.7368}{{\em Phys. Rev.} {\bf D55}
  (1997)  7368--7388},
\href{http://arxiv.org/abs/astro-ph/9611125}{{\tt arXiv:astro-ph/9611125
  [astro-ph]}}.

\bibitem{Paolo}
P.~Creminelli, D.~L. López~Nacir, M.~Simonović, G.~Trevisan, and
  M.~Zaldarriaga, ``{Detecting Primordial $B$-Modes after Planck}'',
  \href{http://dx.doi.org/10.1088/1475-7516/2015/11/031}{{\em JCAP} {\bf 1511}
  (2015) no.~11, 031},
\href{http://arxiv.org/abs/1502.01983}{{\tt arXiv:1502.01983 [astro-ph.CO]}}.

\bibitem{BICEP15}
{\bf BICEP2, Keck Array} Collaboration, P.~A.~R. Ade {\em et al.}, ``{BICEP2 /
  Keck Array VI: Improved Constraints On Cosmology and Foregrounds When Adding
  95 GHz Data From Keck Array}'',
\href{http://arxiv.org/abs/1510.09217}{{\tt arXiv:1510.09217 [astro-ph.CO]}}.

\bibitem{PRISM}
{\bf PRISM} Collaboration, P.~Andre {\em et al.}, ``{PRISM (Polarized Radiation
  Imaging and Spectroscopy Mission): A White Paper on the Ultimate Polarimetric
  Spectro-Imaging of the Microwave and Far-Infrared Sky}'',
\href{http://arxiv.org/abs/1306.2259}{{\tt arXiv:1306.2259 [astro-ph.CO]}}.

\bibitem{21cm}
L.~Book, M.~Kamionkowski, and F.~Schmidt, ``{Lensing of 21-cm Fluctuations by
  Primordial Gravitational Waves}'',
  \href{http://dx.doi.org/10.1103/PhysRevLett.108.211301}{{\em Phys. Rev.
  Lett.} {\bf 108} (2012)  211301},
\href{http://arxiv.org/abs/1112.0567}{{\tt arXiv:1112.0567 [astro-ph.CO]}}.

\bibitem{Chongchitnan:2006pe}
S.~Chongchitnan and G.~Efstathiou, ``{Prospects for direct detection of
  primordial gravitational waves}'',
  \href{http://dx.doi.org/10.1103/PhysRevD.73.083511}{{\em Phys. Rev.} {\bf
  D73} (2006)  083511},
\href{http://arxiv.org/abs/astro-ph/0602594}{{\tt arXiv:astro-ph/0602594
  [astro-ph]}}.

\bibitem{Planck15}
{\bf Planck Collaboration} Collaboration, P.~Ade {\em et al.}, ``{Planck 2015.
  XX. Constraints on inflation}'',
\href{http://arxiv.org/abs/1502.02114}{{\tt arXiv:1502.02114 [astro-ph.CO]}}.

\bibitem{GBRSZ}
J.~Garcia-Bellido, D.~Roest, M.~Scalisi, and I.~Zavala, ``{Lyth bound of
  inflation with a tilt}'',
  \href{http://dx.doi.org/10.1103/PhysRevD.90.123539}{{\em Phys.Rev.} {\bf D90}
  (2014) no.~12, 123539},
\href{http://arxiv.org/abs/1408.6839}{{\tt arXiv:1408.6839 [hep-th]}}.

\bibitem{Lyth}
D.~H. Lyth, ``{What would we learn by detecting a gravitational wave signal in
  the cosmic microwave background anisotropy?}'',
  \href{http://dx.doi.org/10.1103/PhysRevLett.78.1861}{{\em Phys.Rev.Lett.}
  {\bf 78} (1997)  1861--1863},
\href{http://arxiv.org/abs/hep-ph/9606387}{{\tt arXiv:hep-ph/9606387
  [hep-ph]}}.

\bibitem{Lotfi}
L.~Boubekeur and D.~Lyth, ``{Hilltop inflation}'',
  \href{http://dx.doi.org/10.1088/1475-7516/2005/07/010}{{\em JCAP} {\bf 0507}
  (2005)  010},
\href{http://arxiv.org/abs/hep-ph/0502047}{{\tt arXiv:hep-ph/0502047
  [hep-ph]}}.

\bibitem{NI}
K.~Freese, J.~A. Frieman, and A.~V. Olinto, ``{Natural inflation with pseudo -
  Nambu-Goldstone bosons}'',
\href{http://dx.doi.org/10.1103/PhysRevLett.65.3233}{{\em Phys. Rev. Lett.}
  {\bf 65} (1990)  3233--3236}.

\bibitem{AM1}
E.~Silverstein and A.~Westphal, ``{Monodromy in the CMB: Gravity Waves and
  String Inflation}'', \href{http://dx.doi.org/10.1103/PhysRevD.78.106003}{{\em
  Phys. Rev.} {\bf D78} (2008)  106003},
\href{http://arxiv.org/abs/0803.3085}{{\tt arXiv:0803.3085 [hep-th]}}.

\bibitem{AM2}
L.~McAllister, E.~Silverstein, and A.~Westphal, ``{Gravity Waves and Linear
  Inflation from Axion Monodromy}'',
  \href{http://dx.doi.org/10.1103/PhysRevD.82.046003}{{\em Phys. Rev.} {\bf
  D82} (2010)  046003},
\href{http://arxiv.org/abs/0808.0706}{{\tt arXiv:0808.0706 [hep-th]}}.

\bibitem{FI}
M.~Cicoli, C.~P. Burgess, and F.~Quevedo, ``{Fibre Inflation: Observable
  Gravity Waves from IIB String Compactifications}'',
  \href{http://dx.doi.org/10.1088/1475-7516/2009/03/013}{{\em JCAP} {\bf 0903}
  (2009)  013},
\href{http://arxiv.org/abs/0808.0691}{{\tt arXiv:0808.0691 [hep-th]}}.

\bibitem{BM}
D.~Baumann and L.~McAllister, {\em {Inflation and String Theory}}.
\newblock Cambridge University Press, 2015.
\newblock \href{http://arxiv.org/abs/1404.2601}{{\tt arXiv:1404.2601
  [hep-th]}}.
\newblock
\url{https://inspirehep.net/record/1289899/files/arXiv:1404.2601.pdf}.
\newblock

\bibitem{Mazumdar:2014qea}
A.~Mazumdar and P.~Shukla, ``{Model independent bounds on tensor modes and
  stringy parameters from CMB}'',
\href{http://arxiv.org/abs/1411.4636}{{\tt arXiv:1411.4636 [hep-th]}}.

\bibitem{KPZ}
K.~Kooner, S.~Parameswaran, and I.~Zavala, ``{Warping the Weak Gravity
  Conjecture}'',
\href{http://arxiv.org/abs/1509.07049}{{\tt arXiv:1509.07049 [hep-th]}}.

\bibitem{Burgess:2016owb}
C.~P. Burgess, M.~Cicoli, S.~de~Alwis, and F.~Quevedo, ``{Robust Inflation from
  Fibrous Strings}'',
\href{http://arxiv.org/abs/1603.06789}{{\tt arXiv:1603.06789 [hep-th]}}.

\bibitem{Anna}
A.~Achucarro, J.-O. Gong, S.~Hardeman, G.~A. Palma, and S.~P. Patil,
  ``{Features of heavy physics in the CMB power spectrum}'',
  \href{http://dx.doi.org/10.1088/1475-7516/2011/01/030}{{\em JCAP} {\bf 1101}
  (2011)  030},
\href{http://arxiv.org/abs/1010.3693}{{\tt arXiv:1010.3693 [hep-ph]}}.

\bibitem{flattening}
X.~Dong, B.~Horn, E.~Silverstein, and A.~Westphal, ``{Simple exercises to
  flatten your potential}'',
  \href{http://dx.doi.org/10.1103/PhysRevD.84.026011}{{\em Phys. Rev.} {\bf
  D84} (2011)  026011},
\href{http://arxiv.org/abs/1011.4521}{{\tt arXiv:1011.4521 [hep-th]}}.

\bibitem{NGs}
R.~Flauger, M.~Mirbabayi, L.~Senatore, and E.~Silverstein, ``{Productive
  Interactions: heavy particles and non-Gaussianity}'',
\href{http://arxiv.org/abs/1606.00513}{{\tt arXiv:1606.00513 [hep-th]}}.

\bibitem{KL}
R.~Kallosh and A.~D. Linde, ``{Landscape, the scale of SUSY breaking, and
  inflation}'', \href{http://dx.doi.org/10.1088/1126-6708/2004/12/004}{{\em
  JHEP} {\bf 12} (2004)  004},
\href{http://arxiv.org/abs/hep-th/0411011}{{\tt arXiv:hep-th/0411011
  [hep-th]}}.

\bibitem{SasakiStewart}
M.~Sasaki and E.~D. Stewart, ``{A General analytic formula for the spectral
  index of the density perturbations produced during inflation}'',
  \href{http://dx.doi.org/10.1143/PTP.95.71}{{\em Prog. Theor. Phys.} {\bf 95}
  (1996)  71--78},
\href{http://arxiv.org/abs/astro-ph/9507001}{{\tt arXiv:astro-ph/9507001
  [astro-ph]}}.

\bibitem{Wands}
D.~Wands, ``{Multiple field inflation}'',
  \href{http://dx.doi.org/10.1007/978-3-540-74353-8_8}{{\em Lect. Notes Phys.}
  {\bf 738} (2008)  275--304},
\href{http://arxiv.org/abs/astro-ph/0702187}{{\tt arXiv:astro-ph/0702187
  [ASTRO-PH]}}.

\bibitem{kinflation}
C.~Armendariz-Picon, T.~Damour, and V.~F. Mukhanov, ``{k - inflation}'',
  \href{http://dx.doi.org/10.1016/S0370-2693(99)00603-6}{{\em Phys. Lett.} {\bf
  B458} (1999)  209--218},
\href{http://arxiv.org/abs/hep-th/9904075}{{\tt arXiv:hep-th/9904075
  [hep-th]}}.

\bibitem{Garriga:1999vw}
J.~Garriga and V.~F. Mukhanov, ``{Perturbations in k-inflation}'',
  \href{http://dx.doi.org/10.1016/S0370-2693(99)00602-4}{{\em Phys. Lett.} {\bf
  B458} (1999)  219--225},
\href{http://arxiv.org/abs/hep-th/9904176}{{\tt arXiv:hep-th/9904176
  [hep-th]}}.

\end{thebibliography}\endgroup

\bibliographystyle{utphys}

\end{document}